# Ultrahigh Photoresponsivity of Gold Nanodisk Array/CVD MoS$_2$-based Hybrid Phototransistor

*Shyam Narayan Singh Yadav[1], Po-Liang Chen[2], Yu-Chi Yao[3], Yen-Yu Wang[4,5], Der-Hsien Lien[6], Yu-Jung L[4,5], Ya-Ping Hsieh[3], Chang-Hua Liu[2,7], and Ta-Jen Yen[1*]*

[1]Department of Materials Science and Engineering, National Tsing Hua University, Hsinchu 300, Taiwan R.O.C.

[2]Institutes of Photonics Technology, National Tsing Hua University, Hsinchu, 30013, Taiwan

[3]Institute of Atomic and Molecular Sciences, Academia Sinica, Taipei, 10617, Taiwan, R.O.C.

[4]Department of Physics, National Taiwan University, Taipei 106, Taiwan R.O.C.

[5]Research Center for Applied Sciences, Academia Sinica, Taipei 115, Taiwan R.O.C.

[6]Department of Electrical Engineering, National Yang-Ming Chao Tung University, Hsinchu, 300, Taiwan, R.O.C.

[7]Department of Electrical Engineering, National Tsing Hua University, Hsinchu, 30013, Taiwan

Corresponding Author

Email: tjyen@mx.nthu.edu.tw

**Abstract**

Owing to its atomically thin thickness, layer-dependent tunable band gap, flexibility, and CMOS compatibility, $MoS_2$ is a promising candidate for photodetection. However, monolayer $MoS_2$-based photodetectors typically show poor optoelectronic performances, mainly limited by their low optical absorption. In this work, we hybridized CVD-grown monolayer $MoS_2$ with a gold nanodisk (AuND) array to demonstrate a superior visible photodetector through a synergetic effect. It is evident from our experimental results that there is a strong light-matter interaction between AuNDs and monolayer $MoS_2$, which results in better photodetection due to a surface trap state passivation with a longer charge carrier lifetime compared to pristine $MoS_2$. In particular, the AuND/$MoS_2$ system demonstrated a photoresponsivity of $8.7 \times 10^4$ AW$^{-1}$, specific detectivity of $6.9 \times 10^{13}$ Jones, and gain $1.7 \times 10^5$ at 31.84 µWcm$^{-2}$ illumination power density of 632 nm wavelength with an applied voltage of 4.0 V for an AuND/$MoS_2$-based photodetector. To our knowledge, these optoelectronic responses are one order higher than reported results for CVD $MoS_2$-based photodetector in the literature.

# 1. Introduction

Conversion of the optical signal into an electrical signal has become a hot topic of research due to its high demands in various fields e.g. video imaging, biomedical imaging, gas sensing, and motion detection.[1] Although this field becomes mature due to the development of high-performance materials and integration technology, still there is a need for high-performance photodetection devices with complementary metal oxide semiconductor (CMOS) compatibility at sub-10-nm technology nodes.[2] Among the many materials being investigated for potential use in photoactive channels, transition metal dichalcogenides (TMDCs) have gained considerable attention because of their sub-nm thickness and lack of dangling bonds at the dielectric interface.[3, 4] In addition, TMDCs possess thickness-dependent bandgap from 1.0 to 2.5 eV,[5-8] high electron mobility (200 $cm^2V^{-1}S^{-1}$),[9] and excellent current ON/OFF ratio ($\approx 10^8$),[10] thriving to be a promising candidate for photodetector.[11-14] Despite these superior optical and electrical qualities, the monolayer $MoS_2$ is intrinsically limited by its low optical absorption (5%),[15] due to its atomically thin layer. On the other hand, bulk $MoS_2$ possesses an indirect bandgap,[16] which is not suitable for photodetection applications. Therefore, those TMDCs-based photodetectors can only reach the maximum responsivity and gain up to the order of $2.2\times10^3$ $AW^{-1}$ and $10^3$, respectively in the case of CVD-grown $MoS_2$.[17-20]

To further improve the photoresponsivity of a monolayer $MoS_2$-based photodetector, researchers demonstrated various methods e.g. making heterostructures,[21, 22] and hybridizing with waveguides,[23-25] Although heterostructures and waveguides integrated $MoS_2$-based reported work shows considerably high photoresponsivity, enabling a strong light-matter interaction still offer potentials to improve optoelectronic responses.[19, 26-28] By employing a similar mechanism, in this work, we have designed and fabricated an gold nanodisk (AuND) array having localize surface plasmon resonance (LSPR) in the visible region (660 nm) near the band edge of monolayer $MoS_2$ as illustrated by our FDTD simulation and UV-Vis measurement results. Furthermore, we have hybridized a CVD-grown monolayer $MoS_2$ with a AuND array on SiO$_2$/Si substrate and characterized their optical properties. Finally, the optoelectronic properties of the designed phototransistors were probed and observed that the AuND/MoS$_2$ hybrid phototransistor outperformed the MoS$_2$ phototransistor owing to the LSPR induced synergetic enhancement. The maximum responsivity of 8.7× 10$^4$ AW$^{-1}$, detectivity of 6.9× 10$^{13}$ Jones, and gain of 1.7× 10$^5$ at 31.84 µWcm$^{-2}$ illumination power density of 632 nm wavelength with an applied voltage of 4.0 V were achieved. Furthermore, the fabricated phototransistor scrutinized of its broadband response, revealing its capability to exhibit photo response across the entire visible spectrum. Based on our knowledge, the demonstrated optoelectronic performance of our designed AuND/MoS$_2$ phototransistor is enhanced in contrast to the presented results in the literature.[17-20]

## 2. Results and Discussion

Our hybrid photodetector is comprised of CVD-grown monolayer MoS$_2$, beneath an array of AuND, as illustrated in **Figure 1**a. In this hybrid photodetector, MoS$_2$ functions as a photoactive channel layer, and AuND as a plasmonic structure that improves light-matter interaction by enabling LSPR in its vicinity resulting in an enhancement in hot electron injection efficiency[29] as well as photodetector performance.[20, 30-33] The synthesis of monolayer MoS$_2$ and fabrication of AuND array is detailed in the **Experimental Methods**. To design the plasmonic structure a parameter sweep at fixed periodicity for variable AuND diameter was run using finite difference time domain (FDTD) simulation using commercial Lumerical® software (refer to **Experimental Methods** for simulation details) and is shown in Figure 1b. From Figure 1b, it is evident that at a diameter size of 90 nm, the reflectance reaches its maximum at the 660 nm wavelength which is also the PL peak of monolayer MoS$_2$. The electric field distribution at resonance wavelength (660 nm) was plotted and x-z is the view shown in Figure 1c, illustrating that the field is highly confined around the edge of the designed AuND owing to induced LSPR at resonance wavelength. The z- component of the x-z view, x-y view, and normal x-y view are shown in Figure S1 (Supporting Information). The z-component of field distribution in Figures S1a and S1c depict the dipole formation due to incident E-field at AuND.

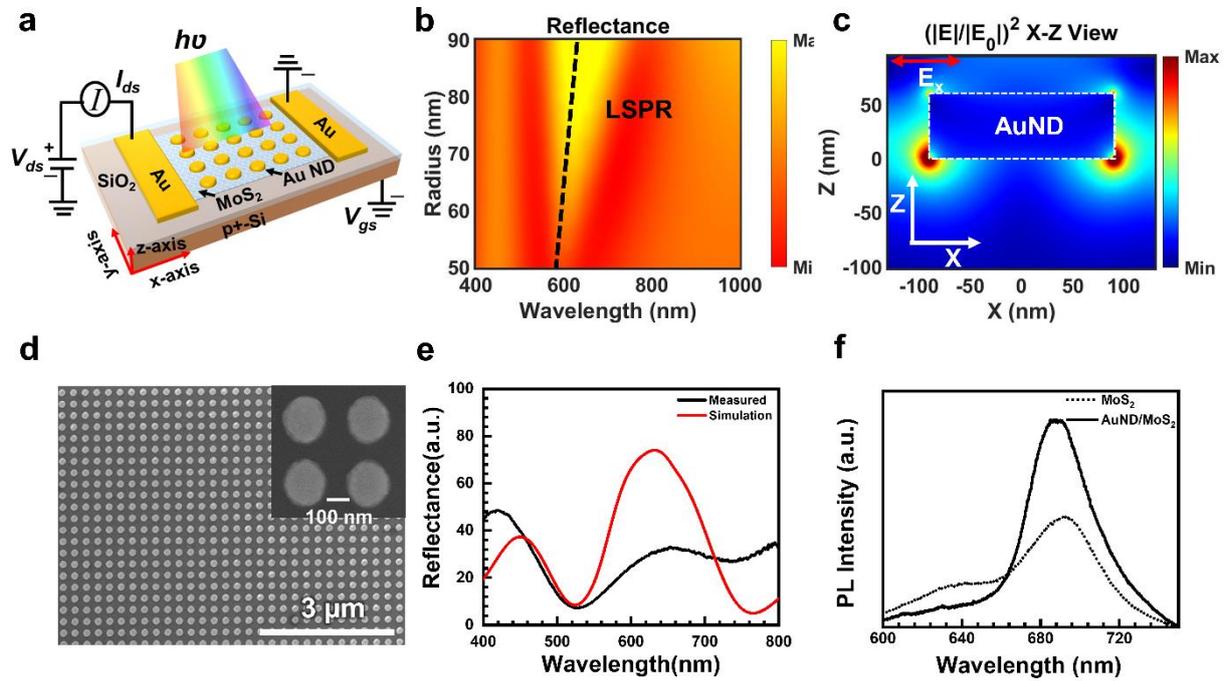

**Figure 1**: (a) The schematic illustration of the AuND array/MoS2 hybrid photodetector device, the inset image shows the fabricated device. (b) Au NDs array parameter optimization. (c) Electric Field E2 distribution in XZ plane (d) SEM image of fabricated Au ND array, the inset image shows enlarge image. (e) Simulated and measured reflectance spectra of designed AuND array. (f) Measurement photoluminescence of pristine MoS2 and AuND/MoS2.

Next, we fabricated the designed AuND array (for fabrication details see **Experimental Methods**). The field emission scanning electron microscopy (FESEM) image of the fabricated AuND array is shown in Figure 1d. It is observed that the size parameters of the fabricated AuND array are well-matched with the designed AuND. Furthermore, the reflectance spectra of the fabricated AuND array were measured using a micro-UV spectrometer and are exhibited in Figure 1e. It is observed that measured reflectance spectra matched well with FDTD-simulated reflectance spectra. In addition, to scrutinize the optical properties of monolayer $MoS_2$ and AuND/$MoS_2$ hybrid structure the UV-Vis absorbance spectra, Raman spectra, and photoluminescence (PL) were measured. The measured raman spectra and absorbance are shown in Figure S2 (Supporting Information). Figure S2a depicts that the two Raman peaks, namely $E^1_{2g}$ and $A_{1g}$ active modes, exhibit nearly 385.64 cm$^{-1}$ and 404.74 cm$^{-1}$, respectively. The difference between these two peaks is calculated and found to be less than 20 cm$^{-1}$, corresponding to monolayer $MoS_2$.[34-36] Furthermore, Figure S2b depicts that the absorbance of $MoS_2$ was enhanced after hybridization with AuND array with resonance peak visible at 660 nm wavelength. This enhancement in absorbance is due to light confinement around the edge of the AuND array and also enlarged light-matter interaction. Next, the measured PL spectra are shown in Figure 1f, illustrating that the PL of the AuND/$MoS_2$ hybrid structure gets enhanced which may be due to strong light-matter interaction and plasmonic-induced resonance energy transfer (PIRET) effect.[26, 37, 38]

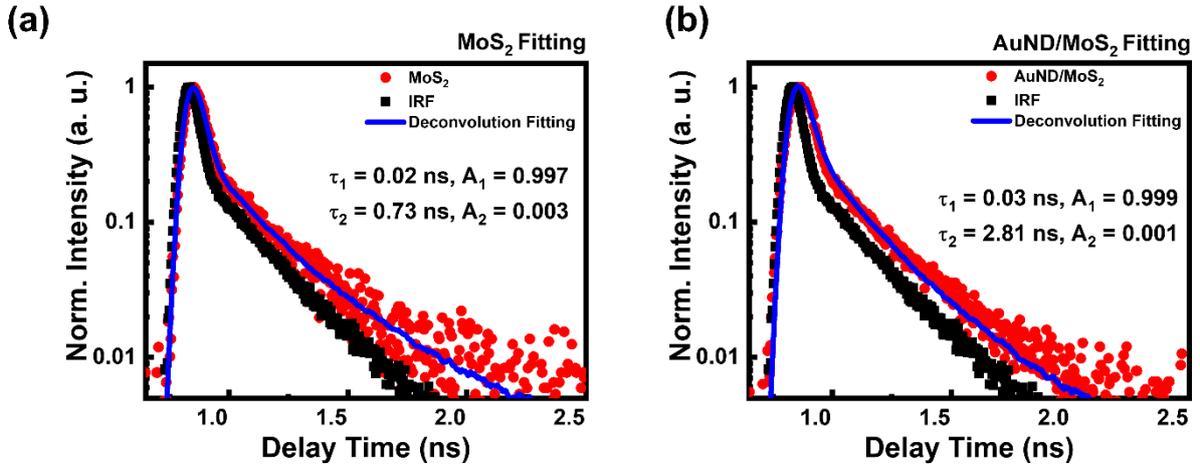

**Figure 2:** Time-resolved photoluminescence (TRPL) measurement **(a)** Fitted TRPL plot for $MoS_2$, and **(c)** Fitted TRPL plot for AuND/$MoS_2$. Here TRPL data was fitted using bi-exponential decay function $f(t) = A_1 exp^{-t/t_1} + A_2 exp^{-t/t_2}$. Here, $A_1$ and $A_2$ are described by amplitude coefficient, and $t_1$ and $t_2$ for a short and long-lifetime corresponding to nonradiative and radiative recombination at surface trap states and band edge, respectively.

To further investigate the effect on excitons dynamics of $MoS_2$ due to the hybridization of the AuND array, we performed the time resolve photoluminescence (TRPL) spectroscopy (for measurement details refer to **Experimental Methods**) as shown in **Figure 2**. The measured TRPL spectra were fitted using biexponential function $f(t) = A_1 exp^{-t/t_1} + A_2 exp^{-t/t_2}$. Here, $A_1$ and $A_2$ are described by amplitude coefficient, and $t_1$ and $t_2$ for short and long-lifetime correspond to nonradiative and radiative recombination at surface trap states and band edge, respectively.[39] The fitted TRPL plot for pristine MoS$_2$ and AuND/MoS$_2$ are exhibited in Figure 2a and Figure 2b, respectively. The calculated parameters after the biexponential fitting are shown in Table S1. (Supporting Information). The short ($t_1$) and long ($t_2$) lifetimes for MoS$_2$ were found to be 0.02 ns and 0.73 ns which is shorter than the 0.03 ns and 2.81 ns of AuNDs and MoS$_2$ hybrid structure. The average lifetime for AuND/MoS$_2$ is again 2.84 ns longer than of MoS$_2$ hybrid structure. This prolonged lifetime of excitons in the AuND/MoS$_2$ hybrid structure is due to trapping state passivation by hot electron injection from AuND to MoS$_2$ and strong resonance coupling between LSPR and exciton, resulting in the suppression of fast exciton recombination. [40]

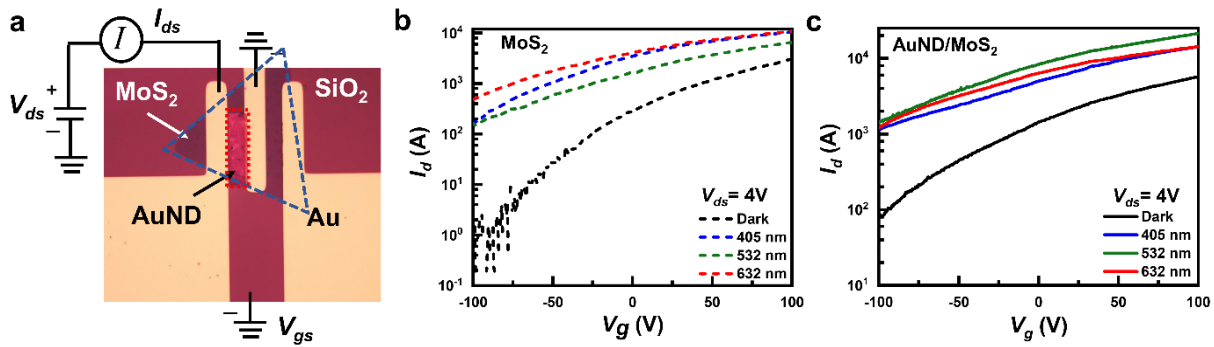

**Figure 3:** (a) The Optical microscope image of fabricated AuND/MoS2 phototransistor with corresponding circute diagram for transfer characteristic measurement. (b) Id−Vg response for pristine MoS2. (c) Id−Vg response of AuND/MoS2 phototransistor.

In addition to the optical characteristics, we further probe the electrical characteristics of MoS$_2$ and AuND/MoS$_2$ hybrid phototransistors. The fabrication of this hybridized phototransistor was carried out using a standard lithography technique (refer to Figure S3, Supporting Information for the device fabrication process). The optical microscope image of the fabricated device with measurement schematic is shown in **Figure 3**a. The transfer characteristics of $MoS_2$ and $AuND/MoS_2$ are illustrated in Figure 3b and Figure 3c, respectively. The normal transfer characteristics are shown in Figure S4 (Supporting Information). The threshold voltage ($V_{th}$), defined by the intercept on $V_{gs}$ with regression fitted line is calculated. The $V_{th}$ for $MoS_2$ phototransistor is −42 V against the $V_{th}$ = −74.8V for the $AuND/MoS_2$ phototransistor, representing the increase in n-type doping after hybridization of AuND with $MoS_2$.[17] This increase in n-type of doping may contribute due to hot electron transfer from AuNDs to $MoS_2$. In addition, the different incident wavelength shows different threshold voltage, indicating a wavelength-dependent electrical characteristic of the $MoS_2$ phototransistor. Further, it is observed that the drain current $I_{ds}$ increased for $AuND/MoS_2$ phototransistors at fixed applied gate voltage. It may be attributed to the confinement of light near the vicinity of AuND and the enhanced population of photoexcited excitons.

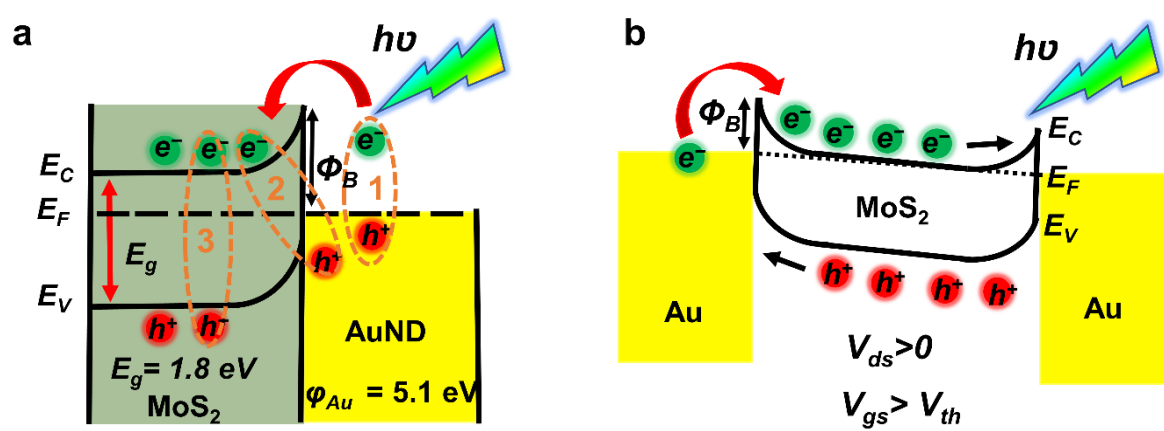

Figure 4: a) Energy band diagram of AuND/MoS2 hybrid structure, representing when MoS2 bring in the proximity of AuND they form a Schottky barrier with height ΦB. A photoinduced plasmonic hot electron from AuND transfer to the conduction band of MoS2, numbers 1, 2, and 3 represent plasmon-induced hot electron transfer, charge carrier transfer, and resonance energy transfer, respectively. b) Charge carrier transfer flow chart of MoS2 under the applied bias Vds> 0 V and gate voltage Vgs > Vth under the incident light.

Next, to explain charge carrier generation and drift mechanism, **Figure 4** illustrates the energy band diagram of the AuND/MoS$_2$ hybrid structure without and with an applied bias. In Figure 4a, a single-layer MoS$_2$ with a Fermi energy of $-$ 4.6 eV,[41, 42] and AuND with a work function of ≈ $-$ 5.1 eV, form a hybrid structure with a Schottky barrier of $\Phi_B$, when they brought into the proximity. Next, when light incident on the hybrid structure three mechanisms take place simultaneously: (1) plasmon-induced hot electron from AuND transfer to MoS$_2$ and passivate the trap state at the interface.[43] (2) LSPR-induced charge carrier transfer directly to the conduction band of MoS$_2$. (3) photoinduced resonance energy transfer generates charge carriers directly in Mos$_2$.[44] In the absence of an external bias, the charge carriers are at their respective energy levels, resulting in zero net currents. However, upon applying an external bias $V_{ds}$ across the electrodes, the photogenerated excitons get dissociated, and the charge carriers from MoS$_2$ begin to drift toward opposite potentials, as depicted in Figure 4b. This leads to a net current flow in contrast to the case when $V_{ds}$ = 0 V.

Next, to evaluated the photo-sensing performance of the hybridized phototransistor, a biased voltage ($V_{ds}$) from 0-4 V was applied across the drain-source contacts to measure the drain current ($I_d$), under the constant back gate-source voltage of $V_{gs}$ = 100 V with different illumination power and wavelengths. First, we compared the photocurrent response with applied bias voltage for pristine MoS$_2$, and AuND/MoS$_2$ with an illumination power at three wavelengths (405nm, 532 nm, and 632 nm) covering the entire visible spectrum of incident light. The photocurrent response for pristine MoS$_2$ and AuND/MoS$_2$ at 632 nm illumination wavelength is shown in Figure 5a, and Figure 5b, respectively. The photocurrents are enhanced by hybridizing AuND with MoS$_2$ at the same illumination power, as expected. Such a result stems from greater optical absorption, which is consistent with our optical measurements mentioned in Figure 1f & Figure S2b (Supporting Information). The photocurrent response with 405 nm and 532 nm laser illumination on MoS$_2$ and AuND/MoS$_2$ was also measured under the same condition and is shown in Figure S6 and Figure S7 (Supporting Information). The photocurrent $(I_{photo})$ was calculated using $(I_{ph} = I_{light} - I_{dark})$. We observed that the device has a better photo response at 632 nm laser compared to that at 405 nm and 532 nm laser illumination. This better performance is due to high light harvesting at the resonance wavelength (exciton-plasmon coupling), consistent with UV-Visible spectra in Figure S2b.

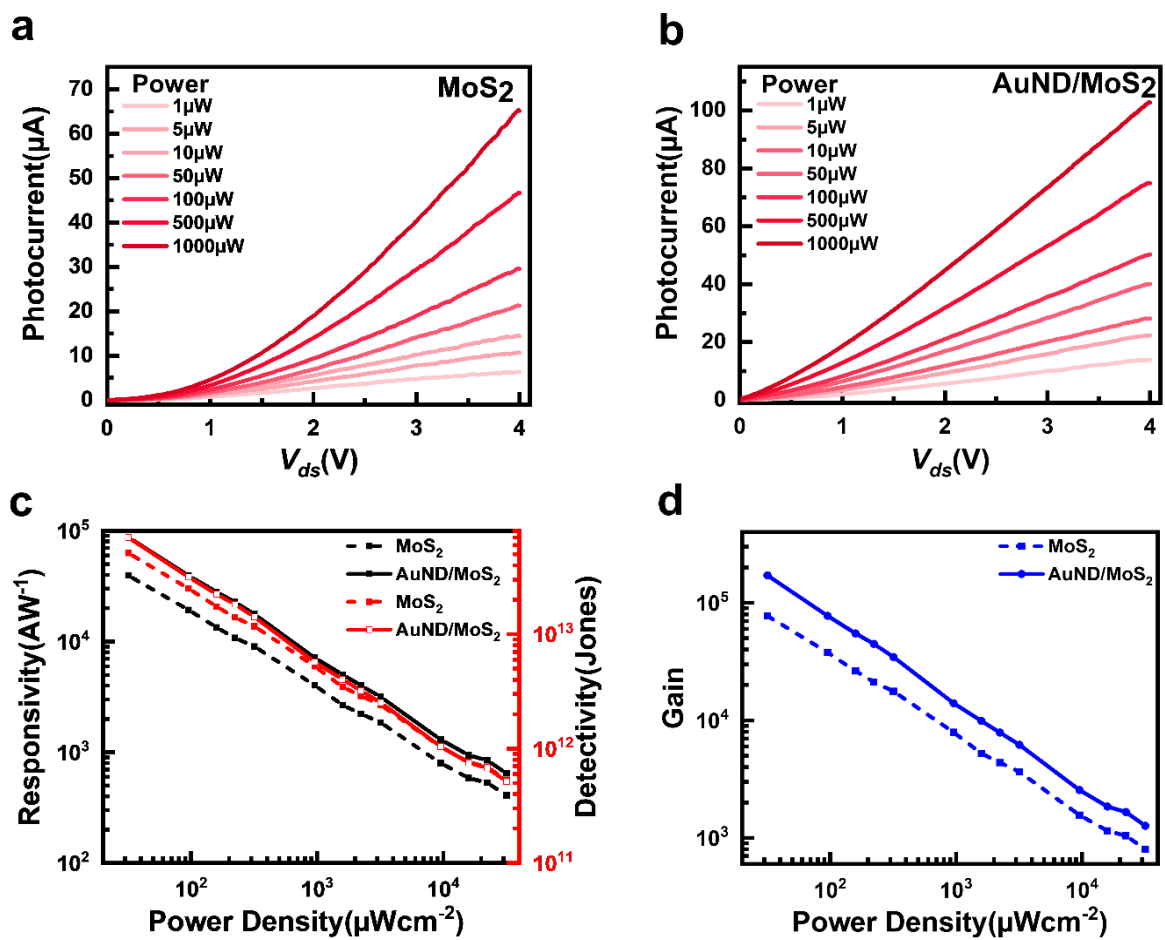

Figure 5: Optoelectrical Characterization: a,b) Photocurrent with applied bias Vds at different illumination power for MoS2 and AuND/MoS2 phototransistors, respectively. c) responsivity and detectivity, d) gain with respect to the incident power density for MoS2 AuND/MoS2 phototransistor at applied bias voltage Vds = 4V, of the incident wavelength of 632nm.

Second, we further studied the optoelectronic characteristics of $MoS_2$, and AuND/$MoS_2$ at the resonance wavelength in this work. The other two wavelengths are also provided in Figure S6c-d and Figure S7c-d (Supporting Information). As shown in Figures 5a and Figure 5b, the photocurrent increases with stronger illumination power, which can be interpreted by the greater population of photoexcited excitons. Based on the measured power-dependent photocurrents, we can characterize three key performance parameters- photoresponsivity (*R*) and specific detectivity (*D*\*), and *Gain*, by the following equations,[45-47]

$$R = \frac{I_{ph}}{P}; \qquad (1)$$

$$D^* = \frac{R_\lambda}{\sqrt{\frac{2qI_{dark}}{A}}}; \qquad (2)$$

$$D^* = \frac{\sqrt{AB}}{NEP}; \qquad (3)$$

$$Gain = \frac{(hcR_\lambda)}{q\lambda}. \qquad (4)$$

where $I_{ph}$, $P$, $R_\lambda$, $q$, $I_{dark}$, $A$, $B$, *NEP*, $h$, $c$, and $\lambda$ denote photocurrent ($I_{ph} = I_{light} - I_{dark}$), illumination power, photoresponsivity at a wavelength of $\lambda$, charge, dark current, active area of the device, bandwidth, noise equivalent power, Planck constant, speed of the light, and wavelength, respectively.

$$NEP = \frac{I_N}{R}; \qquad (5)$$

Here, $I_N$ is the noise current, and *R* is the responsivity of the photodetector. Further, $I_N$ is defined as $I^2_N = 2qI_D B$, where, $I_D$ is the dark current, and *B* is the bandwidth.

The calculated *R*, *D*\*, and *Gain* for both $MoS_2$ and $AuND/MoS_2$ are shown in Figure 5c−d. The *R*, *D*, and *Gain* for other wavelengths are also shown in Figure S6c−d and Figure S6c−d (Supporting Information). Once again, among the two photodetectors, $AuND/MoS_2$ exhibits the best figure of merit (FOM, i.e., *R*, *D*\* and *Gain*) For example, the maximum responsivity, specific detectivity, and Gain of $8.7 \times 10^4$ AW$^{-1}$, detectivity of $6.9 \times 10^{13}$ Jones, and gain $1.7 \times 10^5$ at 31.84 µWcm$^{-2}$ illumination power density of 632 nm wavelength with an applied voltage of 4.0 V. Note that it is a reasonable finding that both *R* and *D*\* decrease along with greater illumination power density, as shown in Figure 5c. Such decrease in the *R* and *D*\* can be attributed to saturation in optical absorption, the screening of the field by photoexcited careers, and enhanced career scattering rate.[48-50] A similar trend was also observed in the calculated *Gain* presented in Figure 5d.

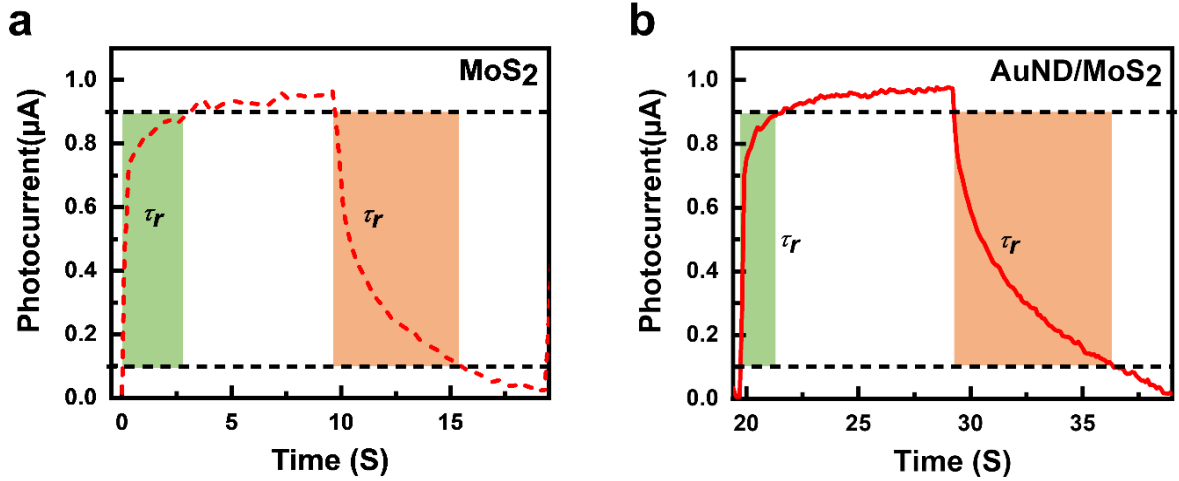

Figure 6: Transient photo response for a) the rise time 3.0 sec and fall time 5.8 sec. for MoS2, b) the rise time 1.7 sec. and fall time 7.2-sec AuND/MoS2 Phototransistor.

Finally, we probed the temporal response of this phototransistor. Both the rise time ($\tau_r$) and the fall time ($\tau_f$) are calculated by the time between 10 % to 90% of the maximum current. As shown in **Figure 6**, the $\tau_r$ and $\tau_f$ for AuND/MoS$_2$ are 1.7 s and 7.2 s, respectively, in contrast to those of pristine MoS$_2$ (3.0 s and 5.8 s). Such a reduction in the response times of this hybrid photodetector originates from the faster carrier excitation caused by the plasmonic effect of AuNDs. Furthermore, to investigate the spectral response of MoS$_2$ and AuND/MoS$_2$ photodetector, the photocurrent measurement with the same applied bias voltage and time at various illumination power of 432 nm and 532 nm wavelength is also performed (see Figure S6 and Figure S7, Supporting information). We observed that our designed AuND/MoS2-based phototransistor shows broadband responses with a considerable optoelectronic performance which is supported by our UV−Vis measurement data. Consequently, we discovered that the designed AuND/MoS$_2$ hybrid photodetector achieved the highest responsivity of $8.7\times10^4$ AW$^{-1}$, detectivity of $6.9\times10^{13}$ Jones, and Gain of $1.7\times10^5$, at the illumination power density of 31.84 µWcm$^{-2}$ of 632 nm wavelength. The demonstrated device performance was also compared with other reported works and is shown in Table 1 (supporting information) which is the highest up-to-date in this class of hybrid photodetectors.

## 3. Conclusions

In conclusion, we designed and fabricated AuND/MoS$_2$ hybrid phototransistor. Firstly, we characterized the optical properties of individual AnuND, MoS$_2$, and their hybrids structures, and observed the synergetic effect from their hybrids. In particular, it is the AuND/MoS$_2$ that outperforms the foremost with the highest optical absorption and PL which is also confirmed by reflectance calculation using the FDTD technique. Secondly, Besides, by using time-resolve photoluminescence to study charge carrier dynamics, such a hybrid structure presented a prolonged charge carrier lifetime (2.84 ns). Finally, the hybrid phototransistor showed the synergy of optoelectronic effects. In particular, the AuND/MoS$_2$ hybrid phototransistor demonstrated the ultrahigh *R*, *D*\*, and *gain* of 8.7× 10$^4$ AW$^{-1}$, 6.9× 10$^{13}$Jones, and 1.7× 10$^5$ at 31.84 µWcm$^{-2}$ illumination power density of 632 nm wavelength with an applied voltage of 4.0 V, and *R* is the highest up to date in the same class of material-based photodetectors. This demonstrated hybrid photodetector shows a high potential to be used for optoelectronic applications.

## 4. Experimental Methods

*Synthesis of monolayer MoS$_2$ by CVD methods and transfer process:* CVD methods were used to produce a large area of monolayer MoS$_2$.[51, 52] The CVD furnace consists of a single heat zone, and a 1-inch quartz tube is used as the reaction chamber. C-plane sapphire is used as substrate. MoO$_3$ powder and H$_2$S are used as the precursor and carrier gas. The first, system is pumped below 10$^{-2}$ torr, where argon is used to purge the chamber. During growth, the furnace is heated to 400 °C in 10 minutes and maintained for 5 minutes, which is followed by the ramp to 900 °C in 15 minutes and maintained for 40 minutes for the growth of monolayer MoS$_2$. During the growth, O$_2$ is introduced starting from 800 °C. After completing, the furnace is subjected to fast cooling at 600 °C. The CVD-grown monolayer MoS$_2$ was then transferred using the wet transfer methods.[53, 54] Firstly, MoS$_2$/c-sapphire is coated with PMMA and dried at 65 °C for 1 hour. Next, the substrate near the interface was etched away with 1M potassium hydroxide (KOH) solution. Following that PMMA supported MoS$_2$ was rinsed three times in DI water to remove the base residues and then transferred to SiO$_2$/Si substrate. The target materials along with the substrate were dried under 65 °C for 3 hours for material adhesion and absorbed water removal. Finally, MoS$_2$ coated substrate was soaked in the acetone for 1 hour to get rid of PMMA. The optical microscope image and atomic force microscope image of transferred MoS$_2$ is shown in Figure S5a with a corresponding thickness of 0.7 nm which confirming CVD-grown MoS$_2$ is a monolayer.

*Fabrication of Au disk array and Electrodes:* Firstly, the gold electrodes on CVD-grown MoS$_2$ were patterned using digital lithography projection (DLP) tools followed by the development, Cr/Au (10/100 nm) deposition using electron beam evaporation, lift-off process. Next, for the fabrication of a gold nanodisk array, an electron beam lithography technique using Elionix ELS 7500-EX was used to expose to pattern Au nanodisk. The development process was done using a mixture of MIBK: IPA (1:3). The Au was deposited using electron beam evaporation methods followed by a liftoff process. The final device was annelid in an argon atmosphere at 150 °C for 2 hours to remove the absorbed moisture and other residual photoresists.

*Numerical simulation:* A commercial Lumerical software package was used to calculate the reflection spectra, and near electric field, distributions using the three-dimensional FDTD approach. A periodic boundary condition was applied with a mesh size of 0.3 nm over the x and y directions and a perfectly matched layer (PML) boundary condition was applied along the z direction. The plane wave source with the normal incident over 400 to 800 nm was used. The refractive index of Silicon, SiO$_2$, and Au (gold) was utilized from the reported data of Johnson and Christy[55].

*Structural characterization:* The structure parameter of fabricated Au nanodisk on MoS$_2$ coated SiO$_2$/Si substrate was examined utilizing field emission scanning electron microscopy (FESEM)

Hitachi SU 8010. The Optical Microscope was used to examine the patterned electrodes on MoS$_2$.

*Optical measurements:* Raman spectra of graphene have been collected using micro-Raman spectroscopy (HORIBA, LabRAM, HR800) with 532 nm solid-state laser excitation. The confocal laser scanning microscope system with a vibration-free closed-cycle cryostat (Attodry 800, attocube) was used to acquire the PL spectra. Through the employment of a 100x objective lens (0.82 NA; attocube), a 405 nm laser source as the excitation was focused to a small area (diameter ~ 1 μm) on the material. The same lens was used to record the PL emission spectra, which were then sent through a 405 nm long-pass filter and into a spectrometer (Andor, SR500i), which was composed of a monochromator and a thermoelectrically cooled CCD camera.

For TRPL measurements, the 400 nm excitation source was generated via SHG (second harmonic generation) in BBO crystal from an 800 nm pulsed laser (Tsunami, Spectra Physics) with a pulsed duration of 100 fs and a repetition rate of 1 kHz. A time-correlated single-photon counting (TCSPC) system (Pico Harp 300, Pico Quant) was employed to record the signal with a resulting time-resolution of 50 ps. Before the photon counting system, a 405 nm long-pass filter was used to remove any remaining light at 400 nm.

For micro UV-Visible spectra a were used to record the optical abortion of Au Nanodisk with MoS$_2$ and only MoS$_2$.

*Optoelectrical measurements:* To quantify the photoresponse of the fabricated devices, photoelectric measurements were conducted using a probe station system (Keithley, 2400 SCS). Lasers emitting at wavelengths of 405 nm, 532 nm, and 632 nm were employed within the visible spectrum to excite the AuND/MoS2 phototransistor.

**Author contributions**

S.N.S. Yadav, and T.J Yen conceived the project. T.J Yen directed the project. S.N.S. Yadav designed the devices using FDTD simulation. S.N.S. Yadav fabricated and characterized it. Y.

C. Yao and Y. P. Hsieh synthesized large-area CVD $MoS_2$ and provide its Raman spectra. Y.-Y. Wang and Y.J. Lung performed the PL, TRPL, and micro UV-visible measurements. P.L. Chen and C.-H. Liu helped to measure optoelectronic characteristics. S.N.S. Yadav process all the data and analyzed it. All authors participated in the preparation of the manuscript and commented on its content.

**Acknowledgments**

This work was financially supported by the "High Entropy Materials Center" from The Featured Areas Research Center Program within the framework of the Higher Education Sprout Project by the Ministry of Education (MOE) and from the Project NSTC 111-2634-F-007-008 - by National Science and Technology Council (NSTC) in Taiwan.

**Conflict of interest**

All authors declare no conflict of interest.

**Table of contents:**

To take advantage of the excitons-plasmon coupling of $MoS_2$ and AuND array, a hybrid structure has been developed. The incorporation of the AuND array in $MoS_2$ enhanced optical characteristics by intensifying the light-matter interaction via localized surface plasmon resonance. In addition, AuNDs also passivate the trap state of $MoS_2$ resulting in a delay in carrier decay. The demonstrated photodetector outperforms this class of photodetector and achieved the highest optoelectronic performance.

*Shyam Narayan Singh Yadav[1], Po-Liang Chen[2], Yu-Chi Yao[3], Yen-Yu Wang[6,7], Der-Hsien Lien[5], Yu-Jung Lu[6,7], Ya-Ping Hsieh[3], Chang-Hua Liu[2,4], and Ta-Jen Yen[1*]*

**Ultrahigh Photoresponsivity of Gold Nanodisk Array/CVD $MoS_2$-based Hybrid Phototransistor**

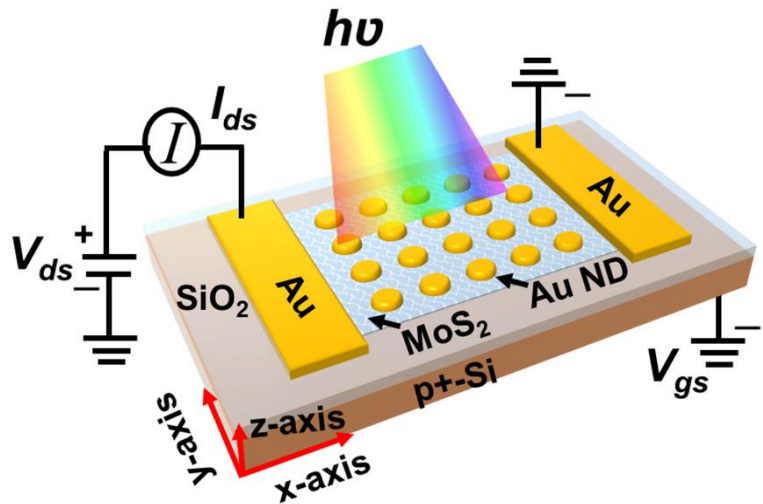

**Supporting Information**

**Ultrahigh Photoresponsivity of Gold Nanodisk Array/CVD MoS$_2$-based Hybrid Phototransistor**

*Shyam Narayan Singh Yadav[1], Po-Liang Chen[2], Yu-Chi Chen[3], Yen-Yu Wang[6,7], Chang Hua Liu[2,4], Ya-Ping Hsieh[3], Der-Hsien Lien[5], Yu-Jung Lu[6,7], and Ta-Jen Yen[1]*

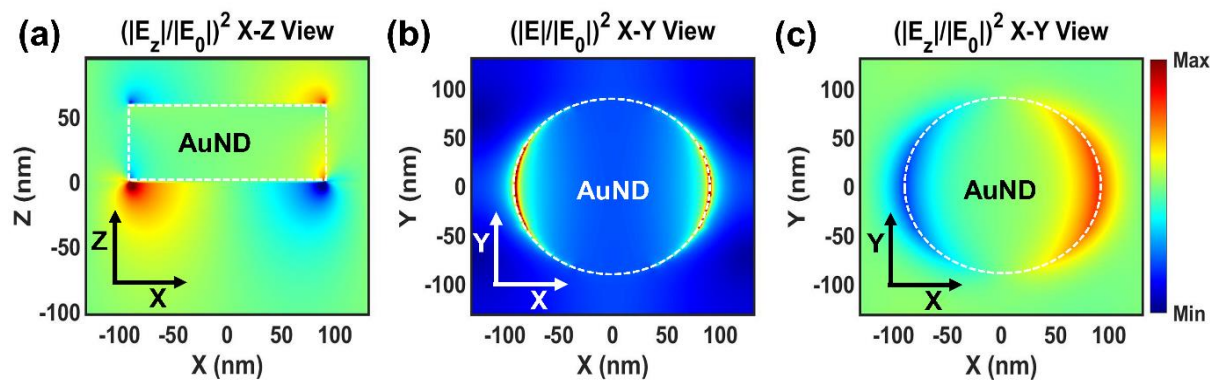

**Figure S1:** Electric field distribution for AuND in (a) X-Z view, (b) X-Y view, and (c) Z component in X-Y view.

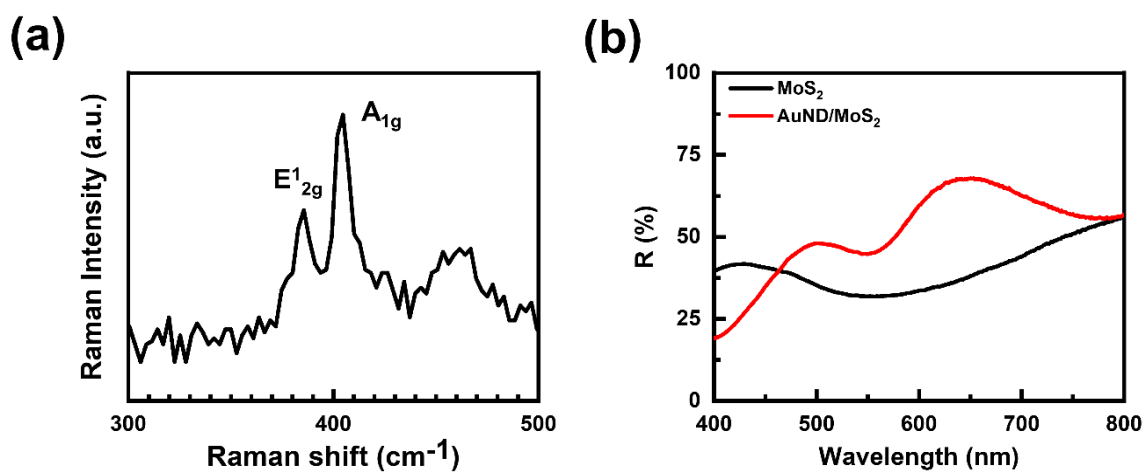

**Figure S2:** (a) Raman spectra of CVD-grown monolayer MoS$_2$, and (b) Reflectance spectra of MoS$_2$ and AuND/MoS$_2$.

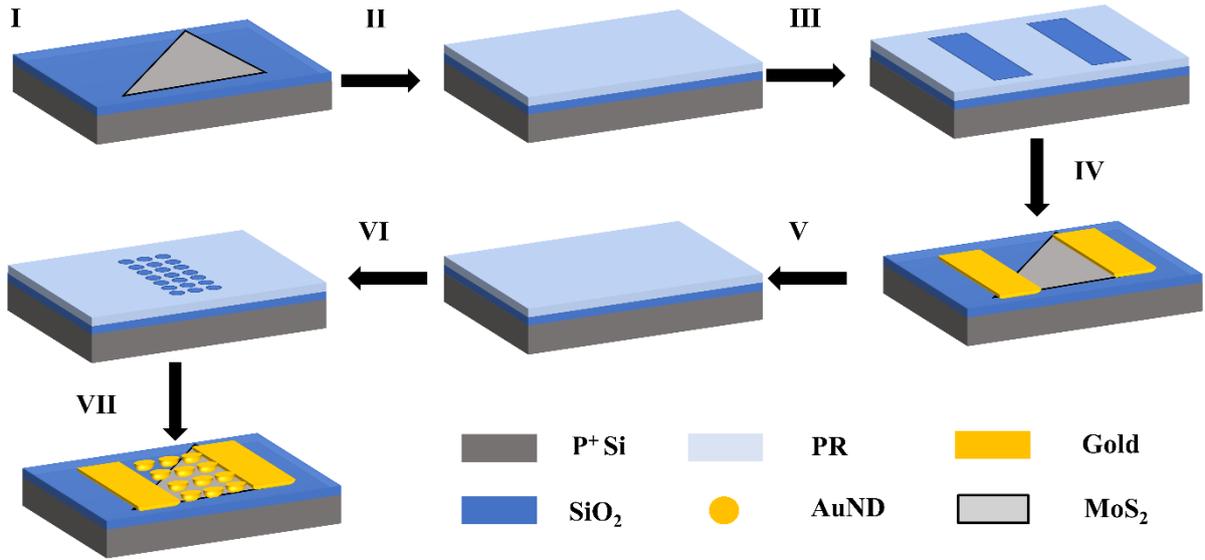

**Figure S3: Device Fabrication Process flow** (I) Transferred MoS2 on SiO$_2$/Si substrate, (II) PR coating, (III) electrode patterned using digital lithography projection (DLP) system and developed, (IV) Au deposition using e gun evaporation system and liftoff (V) PR coating using spin coating system (VI) electron beam lithography for AuND array and development, (VII) Au deposition using e gun evaporation system and liftoff.

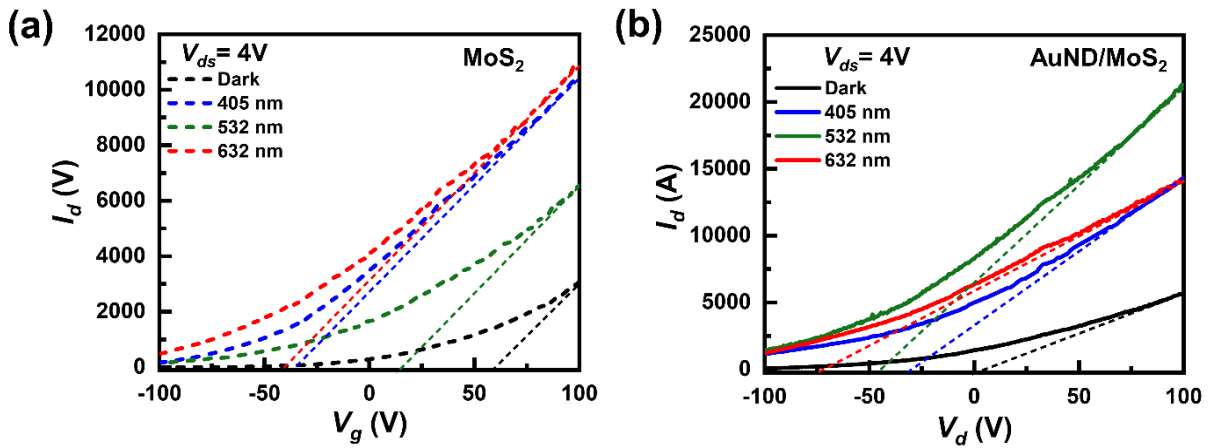

**Figure S4: Transfer characteristic curve of** (a) MoS$_2$, and (b) AuND/MoS$_2$. Here, dotted lines are tangents of the characteristic curve.

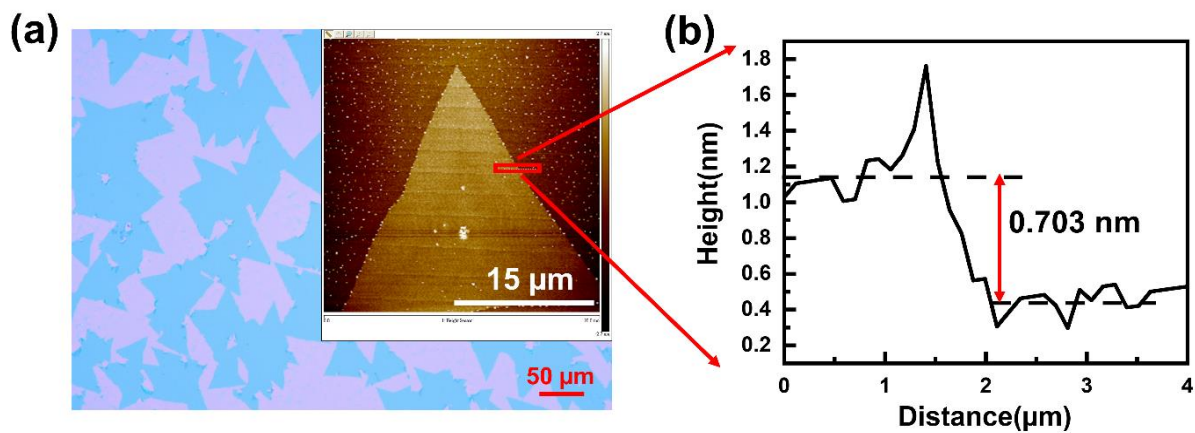

**Figure S5:** (a) Optical Microscope image of CVD-grown MoS$_2$, the inset image is an AFM image of a single flake of MoS$_2$. (b) the height profile of a single flake.

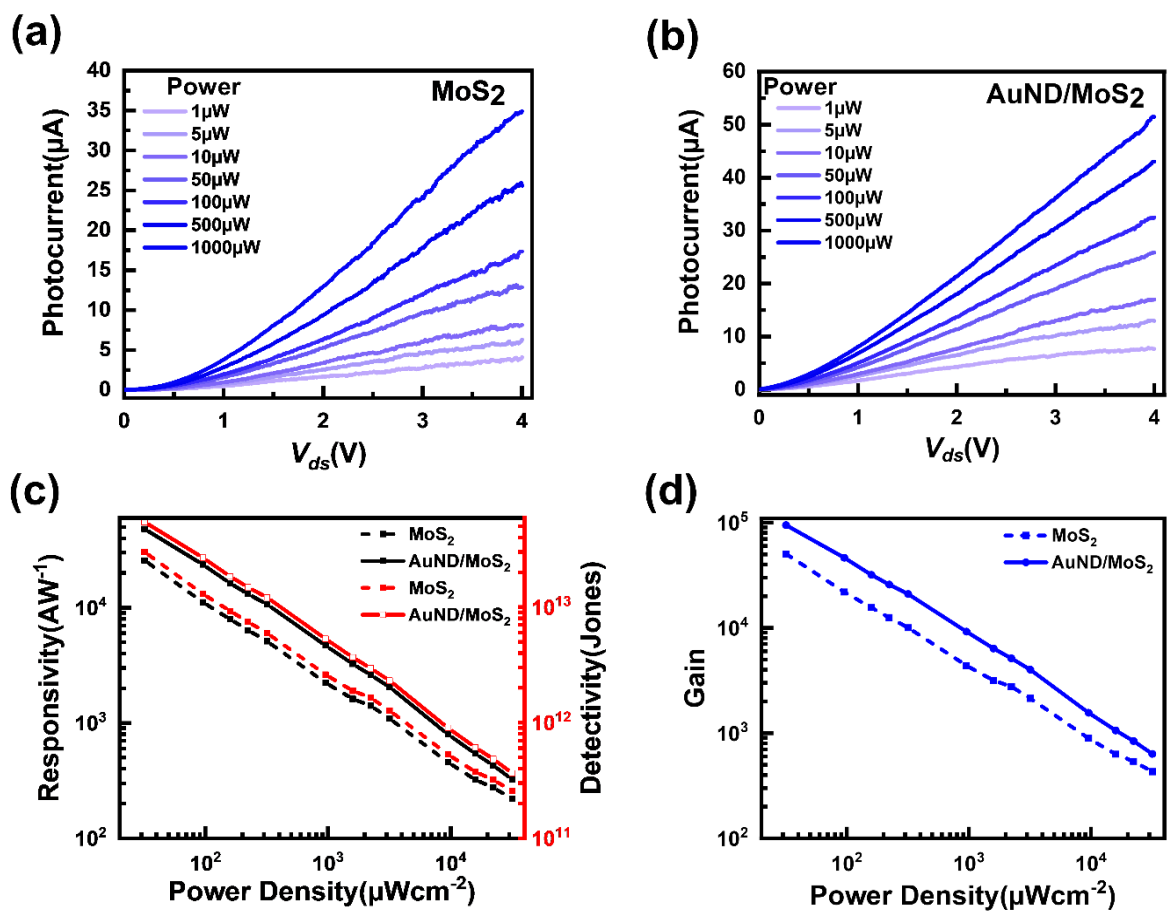

**Figure S6. Optoelectrical Characterization**: **(a,b)** Photocurrent with applied bias $V_{ds}$ at different illumination power for MoS$_2$ and AuND/MoS$_2$ phototransistors, respectively. (c) responsivity and detectivity, (d) gain with respect to the incident power density for MoS$_2$

AuND/MoS$_2$ phototransistor at applied bias voltage $V_{ds}$ = 4V, of the incident wavelength of 405 nm.

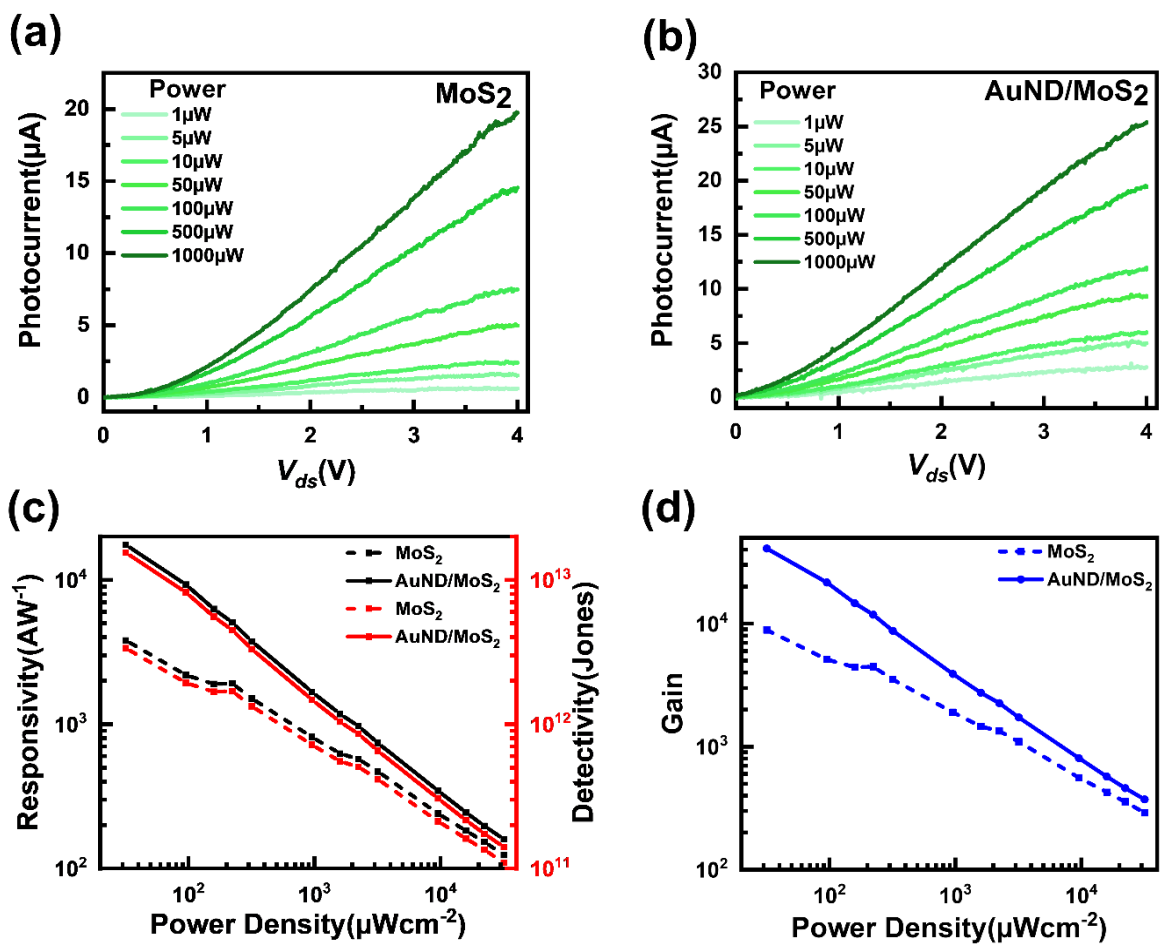

**Figure S7. Optoelectrical Characterization**: **(a,b)** Photocurrent with applied bias $V_{ds}$ at different illumination power for MoS$_2$ and AuND/MoS$_2$ phototransistors, respectively. (c) responsivity and detectivity, (d) gain with respect to the incident power density for MoS$_2$ AuND/MoS$_2$ phototransistor at applied bias voltage $V_{ds}$ = 4V, of the incident wavelength of 532nm.

**Table S1.** The calculated parameters after biexponential fitting of measured TRPL data

| Type of Photodetector | $A_1$ | $t_1$ [ns] | $A_2$ | $t_2$ [ns] |
|---|---|---|---|---|
| MoS$_2$ | 0.997 | 0.02 | 0.003 | 0.73 |
| AuND/MoS$_2$ | 0.999 | 0.03 | 0.001 | 2.81 |

**CS1. The calculation of device parameters (Responsivity (*R*), External quantum efficiency (*EQE*), and specific detectivity (*D\**) for AuND/MoS$_2$-based Phototransistor**

The AuND/MoS2-based phototransistor device parameters were obtained under an illuminating wavelength of 400-700 nm with 1µW (0.15 nW) 31.84713 µWcm$^{-1}$ power and an applied bias voltage of 4 V.

The **photoresponsivity (*R*)** for AuND/MoS$_2$-based phototransistor can be estimated using Equation 1;

$$R = \frac{|I_{photo}|}{P} ; \qquad (1)$$

where $I_{photo}$ is the photocurrent ($I_{photo} = I_{light} - I_{dark}$), and, *P* is the illuminated power to the device [(laser power/laser spot area) × device active area].

$R$ = (13911.64016 nA) / (1000nW*500 µm$^2$/ (3.14*1*10$^6$ µm$^2$) = **8.7 × 10$^4$ AW$^{-1}$**

The specific detectivity (*D\**) of the AuND/MoS$_2$-based phototransistor-based photodetector can be calculated using Equation 2;

$$D^* = \frac{R_\lambda}{\sqrt{\frac{2qI_{dark}}{A}}} ; \qquad (2)$$

where $R_\lambda$, *q*, $I_{dark}$, and *A*, denote the responsivity at a wavelength of *λ*, charge, dark current, and active area of the device, respectively. The active area of the device was measured to be 9490 µm$^2$.

***D\**** = 8.7×10$^4$ AW$^{-1}$× (500 ×10$^{-8}$ cm$^2$)$^{1/2}$/ (2× 1.6×10$^{-19}$ × 24.38913×10$^{-6}$) A)$^{1/2}$ = **6.9×10$^{13}$**

**Jones**

The specific detectivity ($D^*$) of the AuNDs/MoS$_2$-based PD can also be calculated using Equation 3;

$$D^* = \frac{\sqrt{AB}}{NEP};\qquad(3)$$

Where $A$ is the active area, $B$ is the bandwidth which is inversely proportional to response time, and *NEP* is noise equivalent power.

*NEP* can be expressed as Equation 4;

$$NEP = \frac{I_N}{R};\qquad(4)$$

Here, $I_N$ is the noise current, and $R$ is the responsivity of the photodetector. Further, $I_N$ is defined as $I^2_N = 2qI_D B$, where, $I_D$ is the dark current, and $B$ is the bandwidth.

Now $B = 1/\tau = 1/1.7$ sec. = **0.5882 S$^{-1}$**

$I_N = (2\times1.6\times10^{-19} \times 24.38913\times10^{-6}$ A $\times 0.5882$ S$^{-1}$ )$^{1/2}$ = **2.14257×10$^{-12}$ A**

Next, *NEP* can be calculated by $NEP = 2.14257\times10^{-12}/8.7365\times10^4$ AW$^{-1}$ = **2.452×10$^{-17}$ W**

Now, specific detectivity can be calculated as $D^* = (500\times10^{-8}$cm$^2\times 0.5882$ S$^{-1})^{1/2}/ (2.452\times10^{-17}$ W$^{-1}$) = **6.994 ×10$^{13}$ cm Hz$^{1/2}$W$^{-1}$ (Jones)**

The *Gain* for AuND/MoS$_2$-based phototransistor-based -based photodetector can be calculated using Equation 2;

$$Gain = \frac{(hcR_\lambda)}{q\lambda};\qquad(2)$$

where $h$, $c$, $R_\lambda$, $q$, and $\lambda$, indicate the Planck constant, speed of the light, responsivity at a wavelength of $\lambda$, charge, and wavelength of the incident light, respectively.

Here, hc/q = 1240

**Gain** = ((8.7 × 10$^4$ AW$^{-1}$×1240)/632 nm) = **1.7×10$^5$**

**Table S2.** The comparison table for the photoelectric performance of the various reported photodetectors with different configurations.

| Type of Photodetector | Responsivity | Detectivity [Jones] | Gain | Response Time Rise time/Fall time | References |
|---|---|---|---|---|---|
| Au disk/MoS$_2$ | 26.9 AW$^{-1}$ at 420 nm | 6 × 10$^{10}$ | — | | [18] |
| Au Nanoparticle/MoS$_2$ | 2-3 mAW$^{-1}$ at 632 nm | — | — | | [19] |
| CVD MoS$_2$ | 2200 AW$^{-1}$ at 0.4 V, 532 nm | — | 5000 | — | [17] |
| AgND/Exfoliated MoS$_2$ | 2.7× 10$^4$ AW$^{-1}$ at 0.1V, 632 nm | 1.3× 10$^{12}$ | — | — | [20] |
| AuND/MoS$_2$ phototransistor | 8.7× 10$^4$ AW$^{-1}$ at 4V, 632 nm | 6.9× 10$^{13}$ | 1.7× 10$^5$ | 1.7/7.2 | This Work |